\documentclass[11pt]{article}
\usepackage{amsfonts}
\usepackage{amssymb}
\usepackage{amsmath}
\usepackage{amsthm}
\usepackage{epsfig}

\usepackage{color}
\usepackage{alltt}
\usepackage{mathrsfs}
\usepackage{graphicx,ifthen}
\usepackage{subfigure}

\newlength{\bredde}
\def\slash#1{\settowidth{\bredde}{$#1$}\ifmmode\,\raisebox{.15ex}{/}
\hspace*{-\bredde} #1\else$\,\raisebox{.15ex}{/}\hspace*{-\bredde} #1$\fi}
\newcommand{\be}{\begin{equation}}
\newcommand{\ee}{\end{equation}}
\newcommand{\bea}{\begin{eqnarray}}
\newcommand{\eea}{\end{eqnarray}}
\newcommand{\nn}{\nonumber}
\newcommand{\e}{\mbox{e}}
\newcommand{\sect}[1]{\setcounter{equation}{0}\section{#1}}

\def\Tr{{\mbox{Tr}\,}}
\newcommand{\laj}{\lambda^{(j)}}
\newcommand{\La}{\Lambda}
\newcommand{\la}{\lambda}
\newcommand{\tL}{\tilde{L}}
\newcommand{\al}{\alpha}
\newcommand{\ga}{\gamma}

\begin{document}
\topmargin -1.4cm
\oddsidemargin -0.8cm
\evensidemargin -0.8cm
\title{\Large{{\bf
Singular value correlation functions for products of\\ Wishart random matrices}}}

\vspace{1.5cm}

\author{~\\{\sc Gernot Akemann}$^{1,2}$\footnote{akemann@physik.uni-bielefeld.de}, {\sc  Mario Kieburg}$^{1,3}$\footnote{mkieburg@physik.uni-bielefeld.de} and {\sc Lu Wei}$^{4,5}$\footnote{lu.wei@helsinki.fi}
\\~\\~\\
$^1$Department of Physics, Bielefeld University, Postfach 100131, D-33501 Bielefeld, Germany\\~\\
$^2$Laboratori Nazionali di Frascati, Via E. Fermi 40, I-00044 Frascati (Rome), Italy\\~\\
$^3$Department of Physics and Astronomy, SUNY, Stony Brook, New York 11794, USA\\~\\
$^4$Department of Mathematics and Statistics, University of Helsinki, P.O. Box 68, FIN-00014, Finland\\~\\
$^5$School of Electrical Engineering, Aalto University, P.O. Box 13000, FIN-00076 Aalto, Finland}
\date{}
\maketitle
\vfill

\begin{abstract}

We consider the product of $M$ quadratic random matrices with complex elements and no further symmetry, where all matrix elements of each factor have a Gaussian distribution. This
generalises the classical Wishart-Laguerre Gaussian Unitary Ensemble with $M=1$.
In this paper we first compute the joint probability distribution for the singular values of the product matrix when the matrix size $N$ and the number $M$ are fixed but arbitrary. This
leads to a determinantal point process which can be realised in two different ways. First, it can be written as a one-matrix singular value model with a non-standard Jacobian, or second, for $M\geq2$, as a two-matrix singular value model with a set of auxiliary singular values and a weight proportional to the Meijer $G$-function.
For both formulations we determine all singular value correlation functions
in terms of the kernels of biorthogonal polynomials which we explicitly construct. They are given in terms of 
hypergeometric and Meijer $G$-functions, generalising the
Laguerre polynomials for $M=1$.
Our investigation was motivated from applications in telecommunication of multi-layered scattering MIMO channels. We present the ergodic mutual information for finite-$N$ for such a channel model with $M-1$ layers of scatterers as an example.

\end{abstract}

\vfill

\thispagestyle{empty}
\newpage

\renewcommand{\thefootnote}{\arabic{footnote}}
\setcounter{footnote}{0}

\sect{Introduction}\label{intro}

Random matrix theory (RMT) remains a very active field of research, after many decades of work. While originally being conceived in the area of mathematical statistics and nuclear physics, today's applications of RMT
extend beyond the mathematical and physical sciences even in a broad sense, and we refer to \cite{handbook} for a recent overview.

One of the topics in RMT that has caught recent attention is that of products of random matrices. Having one of its original motivations in statistical physics in the description of chaotic and disordered systems \cite{Crisanti}, among  more recent applications are combinatorics \cite{Karol} and telecommunications \cite{Verdu}, which has also been one of our motivations.
In particular, we
consider MIMO (multiple-input and multiple-output) communication networks,
where multi-antenna transceivers are utilised to improve the system
capacity in a rich scattering environment.

Products of matrices loose much of the symmetry of the individual matrices and are generically complex. For simplicity we will consider the individual matrices to be complex, too, with independent Gaussian distributions. The spectral properties of matrix ensembles carry important information. It is encoded in the eigenvalue decomposition as well as in the singular value decomposition. We will focus on the latter here.

A striking property of RMT is its universality, that is the independence of the underlying distribution of the individual matrix elements. It is usually manifest in the limit of large matrix size. However, if we study the local, microscopic behaviour of the spectrum on the scale of the mean level spacing between singular values, it is often vital to have a detailed knowledge of the joint distribution of singular values (or eigenvalues) at hand for finite matrix size. In particular to derive a determinantal or Pfaffian structure of the correlation functions of the random matrix ensemble has proven very useful for universality studies. Some of the most powerful proofs of universality start from the knowledge of orthogonal polynomials of these determinantal or Pfaffian point processes, in order to perform the asymptotic analysis. We refer to \cite{Deift} for a standard reference.

The aim of this work is to provide such a starting point, by deriving the joint probability density function (jpdf) of the singular values of the product matrix at finite matrix size $N$, for a finite product of $M$ matrices. We consider the simplest case of $M$ quadratic $N\times N$ matrices of Wishart type, independently and identically distributed by Gaussians with unit variance. In telecommunications this is also the setting often encountered, with both $N$ and $M$ finite. The singular value distribution of products of complex Wishart matrices is then the setup for the calculation of several information-theoretic quantities.

In previous works  the spectral density of singular values as well as the moments of such product matrices were derived in the macroscopic large-$N$ limit. They use probabilistic methods such as free random variables \cite{BBCC,BG,2002Muller}, field theoretic methods such as planar diagrams \cite{BJLNS}, and inverse Mellin transforms \cite{Karol}. The limit of infinitely many matrices in a product were studied in other works, either for finite-size \cite{Newman1,Newman2,Andy,Forrester} or for infinitely large matrices \cite{Nowak,Janik}, where the problem was mapped to a differential equation.

Very recently the jpdf and its correlation functions of the complex eigenvalues of the matrix ensemble we are considering has been derived for finite $N$ and $M$ \cite{ABu,AStr,Jesper}. Also here the macroscopic large-$N$ density in the complex plane was known previously, see \cite{Burda,goetze,BJLNS} for a collection of works. However the corresponding question about the singular value correlation functions for finite $N$ and $M$ was still open and is addressed in our work.

This article is organised as follows. In section \ref{jpdf} we determine the jpdf of singular values, using two different ways. Section \ref{svcorrel} is devoted to the computation of the correlation functions, by first determining the biorthogonal polynomials associated to this problem in subsection \ref{bOP}, and then the kernel(s) leading to all $k$-point correlators in subsection \ref{Hs}. The spectral density itself is discussed in more detail in subsection \ref{R1lim}, where we compute all its moments and identify the correct rescaling for the macroscopic large-$N$ limit of our results. In particular we compare our results to the known large-$N$ asymptotic spectral density e.g. from \cite{BJLNS}. Section \ref{app} illustrates how our results can help in applications in telecommunications, by computing the ergodic mutual information and comparing it to numerical simulations.
After presenting our concluding remarks and open questions in section \ref{conc} we collect some technical tools in two appendices.

\sect{Joint probability distribution of singular values}\label{jpdf}

We are interested in the singular values of the product $P_M$ of $M$ independent matrices with complex matrix elements of size $N\times N$, $X_j\in {\rm Gl}(N,{\mathbb C})$  for all $j=1,\ldots,M$:
\be
P_M\equiv X_M X_{M-1}\ldots X_1\ .
\label{PMdef}
\ee
Furthermore the matrices $X_j$ are distributed by identical, independent Gaussians,
\be
{\cal P}(X_j)=\exp\left[-\Tr X_j^\dag X_j\right]\ .
\label{Gauss}
\ee
The partition function ${\cal Z}_N^{(M)}$ is then defined as
\be
{\cal Z}_N^{(M)}= C \int \prod_{j=1}^M d[X_j] P(X_j)\ ,
\label{Zdef}
\ee
where $d[X_j]=\prod_{\al,\beta=1}^N d(X_j)_{\al\beta}d(X_j^*)_{\al\beta}$ denotes the flat measure over all independent matrix elements. In this section we compute the jpdf
of the singular values of $P_M$. For $M=1$ this is the well known Wishart-Laguerre (also called chiral Gaussian) Unitary Ensemble, and for most of the following we will thus restrict ourselves to $M>1$. We will not keep track of the normalisation constant $C$ here,
and will only specify it later, once we have changed to singular values.

In the first step we perform the following successive change of variables from $X_j$ to $Y_j$:
\be
Y_1\equiv X_1\ ,\ \ \ \mbox{and}\ \ \ Y_j\equiv X_jY_{j-1}\ \ \ \mbox{for}\ \ \ j=2,\ldots,M\ ,
\label{change}
\ee
e.g. $Y_2=X_2X_1$, $Y_3=X_3X_2X_1$ etc. In the new variables $Y_M=P_M$ is the product matrix we are aiming at.
While the first one, $Y_1=X_1$,  is a trivial relabelling, each subsequent change of variables carries a non-trivial Jacobian given by $1/\det[Y_{j-1}^\dag Y_{j-1}]^N$ for $j=2,\ldots,M$. This can be seen as follows.  Due to $d(Y_j)_{\al\beta}=\sum_{\ga=1}^Nd(X_j)_{\al\ga}(Y_{j-1})_{\ga\beta}$ every column vector of the matrix $Y_j$ acquires a factor $1/\det[Y_{j-1}]$ from the change of variables, and likewise its complex conjugate. Taking into account all $N$ column vectors and their complex conjugates we find the given Jacobian for each $j=2,\ldots,M$, and
we arrive at
\be
{\cal Z}_N^{(M)}= C \int \prod_{i=1}^M d[Y_i] \, \exp\left[-\Tr Y_1^\dag Y_1\right]
\prod_{j=2}^M \frac{1}{\det[Y_{j-1}^\dag Y_{j-1}]^N}
\exp\left[-\Tr Y_{j}^\dag Y_j(Y_{j-1}^\dag Y_{j-1})^{-1}\right]\ .
\label{ZY}
\ee
In writing this we assume that the matrices $X_j$ and thus their products $Y_j$ are invertible\footnote{Our restriction from general complex $N\times N$ matrices to ${\rm Gl}(N,{\mathbb C})$ removes only a set of measure zero.}. In the second step we decompose each matrix $Y_j=V_j\Lambda_j U_j$, $j=1,\ldots,M$ in its angles and singular values, where
$\Lambda_j=\mbox{diag}(\laj_1,\ldots,\lambda_N^{(j)})$ contains the positive singular values $\lambda_a^{(j)}\in{\mathbb R}_+$, $a=1,\ldots,N$, and $U_j\in {\rm U}(N)$ and $V_j\in {\rm U}(N)/{\rm U}(1)^N$ are unitary. The Jacobian resulting from the singular value decomposition of each matrix $Y_j$ is well known and is given in terms of the Vandermonde determinant,
\be
\Delta_N(\Lambda_j)\equiv \prod_{N\geq a>b\geq1}(\laj_a-\laj_b)=\det_{1\leq a,b\leq N}\left[(\laj_a)^{b-1}\right].
\label{Vandermonde}
\ee
Since we encounter the matrices $Y_j$ in the combination $Y_j^\dagger Y_j$, only, the unitary matrices $V_j$ completely drop out which leads to
\bea
{\cal Z}_N^{(M)}&=& C^\prime \int \prod_{i=1}^M \left\{d[V_i] d[U_i] \prod_{a=1}^Nd\la_a^{(i)}\la_a^{(i)}
\right\} \exp\left[-\Tr\Lambda_1^2\right] \Delta_N(\La_1^2)^2\nn\\
&&\times\prod_{j=2}^M \frac{1}{\det[\Lambda_{j-1}^2]^N}
\exp\left[-\Tr U_{j-1}U_{j}^\dag \Lambda_j^2 U_{j} U_{j-1}^\dag\Lambda_{j-1}^{-2}\right]\Delta_N(\La_j^2)^2
\nonumber\\
&=& C^\prime \int \prod_{i=1}^M \left\{d[V_i] d[U_i] \prod_{a=1}^Nd\la_a^{(i)}\la_a^{(i)}
\right\} \exp\left[-\Tr\Lambda_1^2\right] \Delta_N(\La_1^2)^2\nn\\
&&\times\prod_{j=2}^M \frac{1}{\det[\Lambda_{j-1}^2]^N}
\exp\left[-\Tr U_{j}^\dag \Lambda_j^2 U_{j} \Lambda_{j-1}^{-2}\right]\Delta_N(\La_j^2)^2
\ .
\label{ZUVL}
\eea
In the second step we employed the invariance of the Haar measure $d[U_j]$ under $U_j\to U_jU_{j-1}$,
which leads to the decoupling of the integrations over $d[U_1]$ and all the $d[V_j]$ from the rest of the integrals.
The remaining unitary integrations in the last line of eq.~\eqref{ZUVL} can be performed using the so-called Harish-Chandra--Itzykson-Zuber (HCIZ) integral \cite{HC,IZ}\footnote{The Haar measure is normalised such that there is no further proportionality constant on the right hand side.}
\be
\int d[U_j]\exp\left[-\Tr \left(U_{j}^\dag \Lambda_j^2 U_{j} \Lambda_{j-1}^{-2}\right)\right]=\frac{1}{\Delta_N(\La_j^2)\Delta_N(\La_{j-1}^{-2})}
\det_{1\leq a,b\leq N}\left[\exp\left(-\frac{(\la_a^{(j)})^2}{(\lambda_b^{(j-1)})^2}\right)\right]\ .
\label{HCIZ}
\ee
The Vandermonde determinant of inverse powers is proportional to the ordinary one with positive powers due to the following identity:
\be
\Delta_N(\La_j^{-2})=\det_{1\leq a,b\leq N}\left[\frac{1}{(\laj_a)^{2b-2}}\right]
=(-1)^{N(N-1)/2}\frac{ \Delta_N(\La_j^2)}{\det[\Lambda_j^2]^{N-1}}\ .
\label{Vandid}
\ee
This leads to many cancellations in eq. (\ref{ZUVL}), in particular of almost all Vandermonde determinants:
\bea
{\cal Z}_N^{(M)}&=& C" \int_0^\infty \prod_{a=1}^N \left\{ d\la_a^{(M)}\la_a^{(M)}\prod_{j=1}^{M-1}\frac{d\laj_a}{\laj_a}\right\} \exp\left[-\sum_{b=1}^N(\la_b^{(1)})^2\right]\Delta_N(\La_1^2)\Delta_N(\La_M^2)\nn\\
&&\times \prod_{i=2}^M \det_{1\leq c,d\leq N}\left[\exp\left(-\frac{(\la_c^{(i)})^2}{(\lambda_d^{(i-1)})^2}\right)\right]\ .
\label{Zev1}
\eea
We expand the determinant comprising $\lambda^{(1)}_d$ and  $\lambda^{(2)}_c$ in $N!$ terms. Each of these terms yields the same contribution since the permutation involved in the definition of the determinant can be absorbed in the determinant comprising $\lambda^{(2)}_d$ and  $\lambda^{(3)}_c$ due to the antisymmetry of determinants, and a relabelling of the integration variables. Next we expand the determinant comprising $\lambda^{(2)}_d$ and  $\lambda^{(3)}_c$ whose permutations can be absorbed in the determinant comprising $\lambda^{(3)}_d$ and  $\lambda^{(4)}_c$, and so on. This interplay of expansion and absorption of the permutations of the determinants can be continued untill all determinants stemming from the HCIZ integral are replaced by their diagonal part. Note that we do not require any symmetrisation in the variables $\lambda_a^{(M)}$ here. Hence our argument applies to correlation functions of the $\lambda_a^{(M)}$ in the next section \ref{svcorrel}, too, where we integrate the 
joint probability distribution function (jpdf) only over a subset of these singular values.

Almost all remaining multiple integrals can be simplified as follows:
\bea
&&\prod_{a=1}^N\left\{
\int_0^\infty \frac{d\la_a^{(1)}}{\la_a^{(1)}}\,\exp\left[-(\la_a^{(1)})^2\right]
\left(\prod\limits_{j=2}^{M-1}\int_0^\infty\frac{d\la_a^{(j)}}{\la_a^{(j)}}\,\exp\left[-\frac{(\la_a^{(j)})^2}{(\la_a^{(j-1)})^2}\right]\right)\,\exp\left[-\frac{(\la_a^{(M)})^2}{(\la_a^{(M-1)})^2}\right]
\right\} \nn\\
&&\times
\det_{1\leq c,d\leq N}\left[(\la_c^{(1)})^{2d-2}\right]\nn\\
&=&\det_{1\leq c,d\leq N}\left[
\int_0^\infty \frac{d\la_c^{(1)}}{(\la_c^{(1)})^{3-2d}}\exp\left[-(\la_c^{(1)})^2\right]\left(
\prod\limits_{j=2}^{M-1}\int_0^\infty\frac{d\la_c^{(j)}}{\la_c^{(j)}}\,\exp\left[-\frac{(\la_c^{(j)})^2}{(\la_c^{(j-1)})^2}\right]
\right)
\exp\left[-\frac{(\la_c^{(M)})^2}{(\la_c^{(M-1)})^2}\right]
\right]\nn\\
&=&\det_{1\leq c,d\leq N}\left[\frac{1}{2^{M-1}}
G^{M,\,0}_{0,\,M}\left(\mbox{}_{0,\ldots,0,d-1}^{-} \bigg| \, (\la_c^{(M)})^2
\right)\right]\ .
\label{1MMreduct}
\eea
Notice that we have left out the integration over the variables $\la_a^{(M)}$. In the second line of eq.~\eqref{1MMreduct} we have pulled all the integrations into the corresponding rows of the determinants, and in the third line we have used the integral identity~\eqref{Gintid} in the squared singular values, that is derived in appendix~\ref{G-Id}. The special function appearing here is the so-called Meijer $G$-function, see eq.~(\ref{Gint1}) for its definition \cite{Gradshteyn}. The number of zeros in the bottom line of the Meijer $G$-function is $M-1$. Our first main result is thus the following singular value representation of the partition function, after changing to squared singular values $s_a\equiv(\la_a^{(M)})^2$ with $d s_a= 2\la_a^{(M)} d\la_a^{(M)}$, $a=1,\ldots,N$, in eq.~\eqref{Zev1}:
\bea
{\cal Z}_N^{(M)}&=& {C}_{N}^{(M)} \int_0^\infty \prod_{a=1}^N ds_a\
\Delta_N(s)
\det_{1\leq c,d\leq N}\left[
G^{M,\,0}_{0,\,M}\left(\mbox{}_{0,\ldots,0,d-1}^{-} \bigg| \, s_c\right)\right]
= \int_0^\infty \prod_{a=1}^N ds_a {\cal P}_{\text{jpdf}}(s)\ ,
\ \ \ \ \
\label{Zev1MM}\\
{\cal P}_{\text{jpdf}}(s)&\equiv& {C}_{N}^{(M)} \Delta_N(s)
\det_{1\leq c,d\leq N}\left[
G^{M,\,0}_{0,\,M}\left(\mbox{}_{0,\ldots,0,d-1}^{-} \bigg| \, s_c\right)\right]\ ,
\label{jpdf1MM}
\eea
where ${\cal P}_{\text{jpdf}}$ is the jpdf. We will show later that it corresponds to a determinantal point process.
The constant in front of eq.~\eqref{jpdf1MM},
\be
({C}_{N}^{(M)})^{-1}\equiv N!\prod_{a=1}^{N}\Gamma(a)^{M+1} \ ,
\label{Znormalh}
\ee
has been chosen such that the partition function is normalised to unity. This can be seen as follows. Applying the Andr\'eief  integral identity,
\be
\int \prod_{a=1}^N ds_a\ \det_{1\leq c,d\leq N}[\phi_c(s_d)]
\det_{1\leq c,d\leq N}[\psi_c(s_d)]=N!\det_{1\leq c,d\leq N}\left[\int ds\,\phi_c(s)\psi_d(s)\right]\ ,
\label{deB}
\ee
which applies to any two sets of functions $\phi_c$ and $\psi_c$ such that all integrals exist, we obtain  for the partition function
\bea
{\cal Z}_N^{(M)}&=& {C}_{N}^{(M)} N!\,
\det_{1\leq c,d\leq N}\left[\int_0^\infty ds s^{c-1}
G^{M,\,0}_{0,\,M}\left(\mbox{}_{0,\ldots,0,d-1}^{-} \bigg| \, s\right)\right]\nn\\
&=& {C}_{N}^{(M)}N!\,
\det_{1\leq c,d\leq N}\left[ \Gamma(c)^{M-1} \Gamma(c+d-1)\right]
\nn\\
&=& {C}_{N}^{(M)}N!\prod_{c=1}^N\Gamma(c)^{M-1} \prod_{b=1}^N\Gamma(b)^2=1\ .\ \
\label{Zcheck}
\eea
Here we have used another integral identity from the appendix, eq.~\eqref{Gmomid}, and pulled out factors of Gamma-functions from the rows of the determinant. The remaining determinant is nothing but the normalisation of the Wishart-Laguerre ensemble (at $M=1$) which is well-known \cite{Mehta}.

We refer to the result~(\ref{Zev1MM}) as a ``one-matrix" singular value model because it is of the form that would result from the singular value decomposition of a single random matrix, however with a non-standard Jacobian $\neq \Delta_N(s)^2$.
As an easy check we can see that due to
\be
G^{1,\,0}_{0,\,1}\left(\mbox{}_{d-1}^{-} \bigg| \, s_c\right)=s_c^{d-1}\exp[-s_c]\ ,
\ee
the expression in eq. (\ref{Zev1MM}) reduces to the standard Wishart-Laguerre ensemble when setting $M=1$, and taking the exponentials out of the second determinant.

In principle we could now try to compute all singular value $k$-point correlation functions defined as
\be
R_{k}^{(M)}(s_1,\ldots,s_k)\equiv \frac{N!}{(N-k)!}
\int_0^\infty \prod_{a=k+1}^N ds_a\ {\cal P}_{\text{jpdf}}(s)\ .
\label{Rkdef1}
\ee
For example for $k=1$ this gives the spectral density which is normalised to $N$ in our convention following \cite{Mehta},
\be
N=\int ds_1 R_{k=1}^{(M)}(s_1)\ ,
\label{R1norm}
\ee
whereas for $k=N$ we have the jpdf itself, $R_{N}^{(M)}(s_1,\ldots,s_N)=N!{\cal P}_{\text{jpdf}}(s)$.
However, due to the matrix inside the second determinant in eq. (\ref{Zev1MM}) being labelled by indices of the Meijer $G$-function, the computation of the $R_{k}^{(M)}$ is a highly nontrivial task. We postpone this computation to section \ref{svcorrel} using a second "two-matrix" formulation, that is introduced in the next subsection.

\subsection{An alternative jpdf with auxiliary variables}\label{2MMjpdf}

We introduce a formalism that is more convenient to handle when computing correlation functions. Let us step back by considering eq. (\ref{Zev1}), taking for simplicity $M=2$:
\bea
{\cal Z}_N^{(M=2)}&=& C' \int_0^\infty
\prod_{a=1}^N d\la_a^{(2)}\la_a^{(2)}\
\prod_{b=1}^{N}\frac{d\la^{(1)}_b}{\la^{(1)}_b} \exp\left[-(\la_b^{(1)})^2\right]
\ \Delta_N(\La_1^2)\Delta_N(\La_2^2)\nn\\
&&\times\det_{1\leq c,d\leq N}\left[\exp\left(-\frac{(\la_c^{(2)})^2}{(\la_d^{(1)})^2}\right)\right]\ .
\label{Zev2M2}
\eea
This is precisely of the form of a "two-matrix" singular value model (2mm) that results from the singular value decomposition of a two-matrix model, see e.g. in \cite{ADOS}. The advantage is that now we have the {\it standard} form of the Jacobian given by one Vandermonde per set of variables and an additional determinant of a function that couples the two sets of variables. Such a setting can be tackled using the known biorthogonal polynomial technique, as reviewed in \cite{Marco}. In order to apply this technique we have to bring eq. (\ref{Zev1}) into such a form, but for arbitrary values of $M\geq2$. This can be readily achieved by taking the same steps as from eq. (\ref{Zev1}) to eq. (\ref{1MMreduct}), but this time excluding both sets of integrations over the variables $\la_a^{(M)}$ and $\la_a^{(1)}$.

The symmetry argument goes along the same lines replacing determinants by their diagonal parts, see the discussion after eq.~\eqref{Zev1}, but this time we keep the determinant containing the exponential with $\la^{(1)}_d$ and $\la^{(M)}_d$. The integrations are:
\bea
&&\prod_{a=1}^N\prod_{j=2}^{M-1}\left\{
\int_0^\infty\frac{d\la_a^{(j)}}{\la_a^{(j)}}\,\exp\left[-\frac{(\la_a^{(j+1)})^2}{(\la_a^{(j)})^2}\right]
\right\} \det_{1\leq c,d\leq N}\left[\exp\left(-\frac{(\la_c^{(2)})^2}{(\la_d^{(1)})^2}\right)\right]\nn\\
&=&\det_{1\leq c,d\leq N}\left[\prod_{j=2}^{M-1}\left\{
\int_0^\infty\frac{d\la_c^{(j)}}{\la_c^{(j)}}\,\exp\left[-\frac{(\la_c^{(j+1)})^2}{(\la_c^{(j)})^2}\right]
\right\}\exp\left(-\frac{(\la_c^{(2)})^2}{(\la_d^{(1)})^2}\right)
\right]\nn\\
&=&\det_{1\leq c,d\leq N}\left[\frac{1}{2^{M-2}}
G^{M-1,\,0}_{0,\,M-1}\left(\mbox{}_{0,\ldots,0}^{-} \bigg| \, \frac{(\la_c^{(M)})^2}{(\la_d^{(1)})^2}
\right)\right]\ ,
\label{2MMreduct}
\eea
where we have used again the identity (\ref{Gintid}) from appendix \ref{G-Id}, see also \cite{ABu} for a related recurrence relation.
We therefore arrive at our second main result, the following 2mm representation for the jpdf, after changing again to squared singular values $s_a\equiv(\la_a^{(M)})^2$, $t_a\equiv(\la_a^{(1)})^2$, $a=1,\ldots,N$:
\bea
{\cal Z}_N^{(M)}&=& \frac{C_N^{(M)}}{N!} \int_0^\infty
\prod_{a=1}^N ds_a\frac{dt_a}{t_a}\, \e^{-t_a}
\ \Delta_N(s)\Delta_N(t)\det_{1\leq c,d\leq N}\left[
G^{M-1,\,0}_{0,\,M-1}\left(\mbox{}_{0,\ldots,0}^{-} \bigg| \, \frac{s_c}{t_d}\right)
\right]\nn\\
&=&\int_0^\infty \prod_{a=1}^N ds_a\,dt_a {\cal P}^{\text{2mm}}_{\text{jpdf}}(s,t)\ ,
\nn\\
{\cal P}^{\text{2mm}}_{\text{jpdf}}(s,t)&\equiv& \frac{C_N^{(M)}}{N!}
\prod_{a=1}^N t_a^{-1}\e^{-t_a}
\ \Delta_N(s)\Delta_N(t)\det_{1\leq c,d\leq N}\left[
G^{M-1,\,0}_{0,\,M-1}\left(\mbox{}_{0,\ldots,0}^{-} \bigg| \, \frac{s_c}{t_d}\right)\right].
\label{Zev2MM}
\eea
It is normalised to unity as we will check below.

The crucial advantage in comparison to ${\cal P}_{\rm jpdf}$ is the matrix inside the determinant which has indices that label the integration variables, and not the indices of the Meijer $G$-function as in eq. (\ref{Zev1MM}). This 2mm describes the correlations among the singular values $\la_a^{(1)}$ of a single matrix $X_1$ and the singular values $\la_a^{(M)}$ of the entire product matrix $P_M$, to be computed  in the next section \ref{svcorrel}.

As a check for $M=2$ we get back to eq. (\ref{Zev2M2}), using
\be
G^{1,\,0}_{0,\,1}\left(\mbox{}_{0}^{-} \bigg| \, \frac{(\la_c^{(2)})^2}{(\la_d^{(1)})^2}\right)=
\exp\left[-\,\frac{(\la_c^{(2)})^2}{(\la_d^{(1)})^2}\right]\ .
\label{M2check}
\ee
Confirming the normalisation in eq. (\ref{Zev2MM}) is at the same time a check that this representation can be mapped back to eq. (\ref{Zev1MM}) in a different way. Applying once again the Andr\'eief formula (\ref{deB}) to eq. (\ref{Zev2MM}), but this time only to the $t$-integration, we obtain the following:
\bea
&&\int_0^\infty
\prod_{a=1}^N \frac{dt_a}{t_a} \e^{-t_a}
\ \Delta_N(t)\det_{1\leq c,d\leq N}\left[
G^{M-1,\,0}_{0,\,M-1}\left(\mbox{}_{0,\ldots,0}^{-} \bigg| \, \frac{s_c}{t_d}\right)
\right]\nn\\
&=&N!\,\det_{1\leq c,d\leq N}\left[\int_0^\infty dt\,t^{d-2} \e^{-t}
G^{M-1,\,0}_{0,\,M-1}\left(\mbox{}_{0,\ldots,0}^{-} \bigg| \, \frac{s_c}{t}\right)
\right]\nn\\
&=&N!\,\det_{1\leq c,d\leq N}\left[
G^{M,\,0}_{0,\,M}\left(\mbox{}_{0,\ldots,0,d-1}^{-} \bigg| \, {s_c}\right)
\right]\ ,
\eea
upon using the identity (\ref{Gshiftid}) with $m=M-1$ from appendix \ref{G-Id}. This brings us back to the "one-matrix" model representation eq. (\ref{Zev1MM}) in terms of a single set of singular values, with the proper normalisation.

\sect{Singular value correlation functions}\label{svcorrel}

We are now prepared to compute arbitrary $k$-point correlation functions of the singular values.
Rather than using the definition~(\ref{Rkdef1}) we will consider the more general correlation functions of the jpdf in the 2mm representation~(\ref{Zev2MM}):
\be
R_{k,l}^{(M)}(s_1,\ldots,s_k;t_1,\ldots,t_l)\equiv \frac{N!^2}{(N-k)!(N-l)!}
\int_0^\infty \prod_{a=k+1}^N ds_a\ \prod_{b=l+1}^N dt_b\ {\cal P}_{\text{jpdf}}^{\rm 2mm}(s,t)\ .
\label{Rkdef2}
\ee
The $k$-point functions of the singular values of the product matrix $P_M$ in eq. (\ref{Rkdef1}) can be obtained by integrating out all auxiliary variables, or by setting $l=0$: $R_{k,0}^{(M)}(s_1,\ldots,s_k;-)=R_{k}^{(M)}(s_1,\ldots,s_k)$.

Let us introduce the following set of biorthogonal polynomials (bOP) in monic normalisation, $p_i(x)=x^i+\ldots$ and  $q_j(x)=x^j+\ldots$,
\be
\int_0^\infty ds\,dt\, w^{(M)}(s,t)\,p_i^{(M)}(s) q_j^{(M)}(t) = \delta_{ij} h_j^{(M)}\ ,
\label{biOPdef}
\ee
with squared norms $h_j^{(M)}$ and the weight function defined as
\be
w^{(M)}(s,t)\equiv t^{-1}\e^{-t}G^{M-1,\,0}_{0,\,M-1}\left(\mbox{}_{0,\ldots,0}^{-} \bigg| \, \frac{s}{t}\right)\ ,
\label{biwdef}
\ee
for $M>1$. These polynomials are guaranteed to exist following the general theory of bOP that was very recently further developed \cite{Bertola09}, see also \cite{Marco} for a recent review. We will
explicitly construct the bOP for general $M>1$. The general $(k,l)$-point correlation functions eq. (\ref{Rkdef2}) are given in terms of four kernels that are constructed from the kernel of bOP \cite{EM,KieGuh10,Bertola12}:
\be
R_{k,l}^{(M)}(s_1,\ldots,s_k;t_1,\ldots,t_l)=
\det\left[
\begin{array}{cc}
{\underset{1\leq a,b\leq k}{H_{01}(s_a,s_b)}}
& 
\underset{1\leq j\leq l}{\underset{1\leq a\leq k}{H_{00}(s_a,t_j)}}\\
&\\
\underset{1\leq i\leq l}{\underset{1\leq b\leq k}{H_{11}(t_i,s_b)}} & \underset{1\leq i,j\leq l}{H_{10}(t_i,t_j)}
\\
\end{array}
\right]\ .
\label{Rkldet}
\ee
The kernel of bOP is defined as
\be
K_{N}(s,t)\equiv \sum_{j=0}^{N-1}\frac{p_j^{(M)}(s) q_j^{(M)}(t)}{h_j^{(M)}}\ .
\label{Kerdef}
\ee
All four kernels $H_{ab}$ are based on this relation,
\bea
H_{01}(s_a,s_b)&\equiv& \int_0^\infty dt\ K_{N}(s_a,t)t^{-1}\e^{-t}G^{M-1,\,0}_{0,\,M-1}\left(\mbox{}_{0,\ldots,0}^{-} \bigg| \, \frac{s_b}{t}\right)\ ,
\label{H01}\\
H_{00}(s_a,t_j)&\equiv& t_j^{-1}\e^{-t_j}K_{N}(s_a,t_j)\ ,
\label{H00}\\
H_{11}(t_i,s_b)&\equiv&\int_0^\infty ds\int_0^\infty dtK_{N}(s,t)
{t}^{-1}\e^{-t}G^{M-1,\,0}_{0,\,M-1}\left(\mbox{}_{0,\ldots,0}^{-} \bigg| \, \frac{s}{t_i}\right)
G^{M-1,\,0}_{0,\,M-1}\left(\mbox{}_{0,\ldots,0}^{-} \bigg| \, \frac{s_b}{t}\right)
\nn\\
&&-\ G^{M-1,\,0}_{0,\,M-1}\left(\mbox{}_{0,\ldots,0}^{-} \bigg| \, \frac{s_b}{t_i}\right),
\label{H11}\\
H_{10}(t_i,t_j)&\equiv& t_j^{-1}\e^{-t_j}\int_0^\infty ds\,K_{N}(s,t_j)G^{M-1,\,0}_{0,\,M-1}\left(\mbox{}_{0,\ldots,0}^{-} \bigg| \, \frac{s}{t_i}\right)\ .
\label{H10}
\eea
Note that eq. (\ref{Rkldet}) implies that both the one- and two-matrix model jpdf represent determinantal point processes, i.e. for the former eq. (\ref{jpdf1MM}) becomes
\be
{\cal P}_{\text{jpdf}}(s)=\frac{1}{N!}\det_{1\leq a,b\leq N}\left[H_{01}(s_a,s_b) \right]\ .
\label{1MMdet}
\ee

\subsection{The biorthogonal polynomials}\label{bOP}

In order to compute the bOP let us first determine the bimoment matrix
\bea
I_{ij}&\equiv&\int_0^\infty ds\int_0^\infty\,dt\, w^{(M)}(s,t)s^i t^j\nn\\
&=&\int_0^\infty dt t^j\e^{-t}
\int_0^\infty \frac{ds}{t} s^i G^{M-1,\,0}_{0,\,M-1}\left(\mbox{}_{0,\ldots,0}^{-} \bigg| \, \frac{s}{t}\right)\nn\\
&=&\int_0^\infty dt\,t^{j+i}\e^{-t}(i!)^{M-1}\nn\\
&=&(i+j)!(i!)^{M-1}\ ,
\label{Idef}
\eea
which follows again from an identity for Meijer $G$-functions, see eq. (\ref{Gmomid}) appendix \ref{G-Id}. The bOP as well as their norms are determined by this bimoment matrix (see e.g. \cite{Bertola09}):
\bea
p_n^{(M)}(s)&=&\frac{1}{D_n^{(M)}}\det\left[
\begin{array}{cccc}
I_{00}&I_{10}&\ldots&I_{n0}\\
I_{01}&I_{11}&\ldots&I_{n1}\\
\vdots& \vdots&\vdots &\vdots\\
I_{0n-1}&I_{1n-1}&\ldots&I_{nn-1}\\
1&s&\ldots&s^n\\
\end{array}
\right],
\label{pndef}\\
q_n^{(M)}(t)&=&\frac{1}{D_n^{(M)}}\det\left[
\begin{array}{ccccc}
I_{00}&I_{10}&\ldots&I_{n-10}&1\\
I_{01}&I_{11}&\ldots&I_{n-11}&t\\
\vdots& \vdots&\vdots &\vdots&\vdots\\
I_{0n}&I_{1n}&\ldots&I_{n-1n}&t^n\\
\end{array}
\right],
\label{qndef}
\eea
where
\bea
D_n^{(M)}&\equiv& \det_{0\leq i,j\leq n-1}[I_{ij}]=\prod_{i=0}^{n-1}(i!)^{M-1}\det_{0\leq i,j\leq n-1}[(i+j)!]\ ,
\label{Dndef}\\
h_n^{(M)}&=& D_{n+1}^{(M)}/D_n^{(M)}\ .
\label{norm}
\eea
In order to have more explicit expressions it is instructive to compare these equations with the standard Laguerre polynomials $L_n(x)$. We need them in monic normalisation denoted by $\tL_n(x)$:
\be
\tL_n(x)\equiv(-1)^nn!\,L_n(x)=\sum_{k=0}^n\frac{(-1)^{n-k}}{(n-k)!}\left(\frac{n!}{k!}\right)^2x^k\ ,
\label{Lagdef}
\ee
with squared norms
\be
\int_0^\infty dx\, \e^{-x}\tL_n(x)\tL_m(x)=\delta_{nm}(n!)^2\equiv\delta_{nm}\,h_n^{(M=1)} \ ,
\label{normLag}
\ee
and a symmetric bimoment matrix
\be
I_{ij}\big|_{M=1}\equiv\int_0^\infty dt\,t^{i+j}\,\e^{-t}=(i+j)!\ ,\ \  \tilde{L}_n(x)=\frac{1}{D_n^{(M=1)}}\det\left[\begin{array}{ccc} I_{00}\big|_{M=1} & \ldots & I_{0 n}\big|_{M=1} \\ \vdots & \vdots & \vdots \\ I_{0 n-1}\big|_{M=1} & \ldots & I_{n n-1}\big|_{M=1} \\ 1 & \ldots & x^n \end{array}\right].
\label{IM1def}
\ee
The former equals eqs. (\ref{Idef}) with $M=1$. As can be seen the monic Laguerre polynomials also have a determinant representation from the Gram-Schmidt procedure, which is exactly the one in eq. (\ref{pndef}) at $M=1$ (or eq. (\ref{qndef}) as they become equal then).

From the comparison of  eqs. (\ref{pndef}) and \eqref{qndef} to eq.~\eqref{IM1def} we can read off the following. For the $q_n(t)$ we take out the common factors $(i!)^{M-1}$ from the first $n$ columns of the determinant, $i=0,1,\ldots,n-1$, with the remaining determinant being identical to that of the monic Laguerre polynomials. The determinant of the bimoment matrix~(\ref{Dndef}) is already written to be proportional to the corresponding one of the Laguerre ensemble. We thus have
\be
q_n^{(M)}(t)=\frac{\prod_{i=0}^{n-1}(i!)^{M-1}}{\prod_{i=0}^{n-1}(i!)^{M-1}}\tL_n(t)=\tL_n(t)\ ,
\label{qnres}
\ee
for all values of $M$.
Likewise we can read off the squared norms by comparing them to the Laguerre case eq. (\ref{normLag}):
\be
h_n^{(M)}=\frac{\prod_{i=0}^{n}(i!)^{M-1}}{\prod_{i=0}^{n-1}(i!)^{M-1}}h_n^{(M=1)}=(n!)^{M+1}\ .
\label{normfin}
\ee
This equation is formally redundant for $M=1$.

For the polynomials $p_n(s)$ the case is slightly more complicated. Also for these polynomials we can take out common factors from all $n+1$ columns, however this will modify the arguments in the last row of the determinant in the numerator: $s^i\to s^i/(i!)^{M-1}$. Expanding with respect to the last row we get a polynomial with the same coefficients as the monic Laguerre polynomials, but now with $s^i/(i!)^{M-1}$ instead of the monomial $s^i$ alone, resulting into
\bea
p_n^{(M)}(s)&=&\frac{\prod_{i=0}^{n}(i!)^{M-1}}{\prod_{i=0}^{n-1}(i!)^{M-1}}
\sum_{k=0}^n\frac{(-1)^{n-k}}{(n-k)!}\left(\frac{n!}{k!}\right)^2\frac{s^k}{(k!)^{M-1}}\nn\\
&=&\sum_{k=0}^n\frac{(-1)^{n-k}}{(n-k)!}\left(\frac{n!}{k!}\right)^{M+1}s^k\nn\\
&=&(-1)^n(n!)^{M}\ ~_1F_{M}(-n;1,\ldots,1;s).
\label{pnres}
\eea
This function can be interpreted as a generalisation of the monic Laguerre polynomials reobtained when setting $M=1$, see eq. (\ref{Lagdef}). Moreover we could express it in terms of the generalised hypergeometric function $~_1F_{M}$ which has $M$ arguments equal to $1$ in the second set of its indices.

The kernel~(\ref{Kerdef}) is now completely determined. We proceed by computing the various kernels~(\ref{H01}), (\ref{H11}) and (\ref{H10}) by integrating the kernel of bOP.

\subsection{The kernels and all correlation functions}\label{Hs}

We start with the kernel $H_{01}$ that is relevant for the correlation functions of all singular values $s_a=(\la_a^{(M)})^2$.
We have
\bea
H_{01}(s_a,s_b)&=&
\sum_{j=0}^{N-1}\frac{1}{h_j^{(M)}}p_j^{(M)}(s_a)\int_0^\infty dt'\tL_j(t')
{t'}^{-1}\e^{-t'}
G^{M-1,\,0}_{0,\,M-1}\left(\mbox{}_{0,\ldots,0}^{-} \bigg| \, \frac{s_b}{t'}\right)\nn\\
&\equiv& \sum_{j=0}^{N-1}\frac{1}{h_j^{(M)}}p_j^{(M)}(s_a)\chi_j^{(M)}(s_b) \ ,
\label{H01pchi}
\eea
where we introduce the following integral transform
\bea
\chi_j^{(M)}(s_b)&\equiv&\int_0^\infty dt'\tL_j(t')
{t'}^{-1}\e^{-t'}
G^{M-1,\,0}_{0,\,M-1}\left(\mbox{}_{0,\ldots,0}^{-} \bigg| \, \frac{s_b}{t'}\right)\nn\\
&=&\sum_{i=0}^j\frac{(-1)^{j-i}}{(j-i)!}\left(\frac{j!}{i!}\right)^2
\int_0^\infty {dt'}(t')^{i-1}\,\e^{-t'}
G^{M-1,\,0}_{0,\,M-1}\left(\mbox{}_{0,\ldots,0}^{-} \bigg| \, \frac{s_b}{t'}\right)\nn\\
&=&\sum_{i=0}^j\frac{(-1)^{j-i}}{(j-i)!}\left(\frac{j!}{i!}\right)^2
G^{M,\,0}_{0,\,M}\left(\mbox{}_{0,\ldots,0,i}^{-} \bigg| \, {s_b}\right)\ ,
\label{chijdef}
\eea
upon using the identity eq. (\ref{Gshiftid}) for $d-1=i$ and $m=M-1$. A more compact expression of $\chi_j$ can be found by combining the Rodrigues formula
\bea
\tL_j(t')=e^{t'}\left(-\frac{d}{dt'}\right)^j\left({t'}^j e^{-t'}\right)\ ,
\label{Rod}
\eea
and the identity~\eqref{a12}.  We substitute $t^\prime\to s_b/t^\prime$ in eq.~\eqref{chijdef} and  express the derivative in eq.~\eqref{Rod} as a derivative in $s_b$. The integration over $t^\prime$ yields
\bea
\chi_j^{(M)}(s_b)&=&\left(-\frac{d}{ds_b}\right)^j\left(s_b^j G^{M,\,0}_{0,\,M}\left(\mbox{}_{0,\ldots,0}^{-} \bigg| \, s_b\right)\right)=(-1)^jG^{M,\,1}_{1,\,M+1}\left(\mbox{}_{0,\ldots,0}^{-j} \bigg| \, s_b\right).
\label{chijalt}
\eea
The second equality can be obtained by applying the definition~\eqref{Gint1} of Meijer's G-function. We can thus write down the full answer for all $k$-point correlation functions of the singular values:
\bea
R_{k}^{(M)}(s_1,\ldots,s_k)
&=&
R_{k,0}^{(M)}(s_1,\ldots,s_k;-)
=\det_{1\leq a,b\leq k}\left[H_{01}(s_a,s_b)
\right]\nn\\
&=&\det_{1\leq a,b\leq k}\left[
\sum_{j=0}^{N-1}\frac{1}{j!} ~_1F_{M}(-j;1,\ldots,1;s_a)
G^{M,\,1}_{1,\,M+1}\left(\mbox{}_{0,\ldots,0}^{-j} \bigg| \, s_b\right)
\right].\ \ \ \ \
\label{Rkfinal}
\eea
The simplest example is the density or 1-point correlation function of singular values which is given by
\bea
R_{1}^{(M)}(s)&=&H_{01}(s,s)=\sum_{j=0}^{N-1}
\frac{1}{j!} ~_1F_{M}(-j;1,\ldots,1;s)
G^{M,\,1}_{1,\,M+1}\left(\mbox{}_{0,\ldots,0}^{-j} \bigg| \, s\right)
\ .
\label{density}
\eea
This example will be further discussed in the next subsection \ref{R1lim}.

The next kernel $H_{00}$ is readily given from its definition eq. (\ref{H00}) together with eqs. (\ref{pnres}),
(\ref{qnres}) and (\ref{normfin}). We therefore turn to $H_{10}$ from eq. (\ref{H10})
\bea
H_{10}(t_i,t_j)&=&{t_j}^{-1}\e^{-t_j}
\sum_{l=0}^{N-1}\frac{1}{h_l^{(M)}}
\left(\int_0^\infty ds\ p_l^{(M)}(s)
G^{M-1,\,0}_{0,\,M-1}\left(\mbox{}_{0,\ldots,0}^{-} \bigg| \, \frac{s}{t_i}\right)
\right)\tL_l(t_j)
\nn\\
&\equiv& {t_j}^{-1}\e^{-t_j}
\sum_{l=0}^{N-1}\frac{1}{h_l^{(M)}}\psi_l^{(M)}(t_i)\tL_l(t_j) \ ,
\label{H10psiq}
\eea
where we define the following integral transform
\bea
\psi_l^{(M)}(t)&\equiv&
\int_0^\infty ds\ p_l^{(M)}(s)
G^{M-1,\,0}_{0,\,M-1}\left(\mbox{}_{0,\ldots,0}^{-} \bigg| \, \frac{s}{t_i}\right)
\nn\\
&=&\sum_{i=0}^l\frac{(-1)^{l-i}}{(l-i)!}\left(\frac{l!}{i!}\right)^{M+1}
\int_0^\infty {ds}\ s^{i}\,
G^{M-1,\,0}_{0,\,M-1}\left(\mbox{}_{0,\ldots,0}^{-} \bigg| \, \frac{s}{t}\right)\nn\\
&=& \sum_{i=0}^l\frac{(-1)^{l-i}}{(l-i)!}\left(\frac{l!}{i!}\right)^{M+1} t^{i+1} (i!)^{M-1}\nn\\
&=&(l!)^{M-1}t\tL_l(t)\ .
\label{psidef}
\eea
In the second step we have used the identity~(\ref{Gmomid}), which leads us back to the standard Laguerre polynomials. Taking into account the normalisation~(\ref{normfin}) we arrive at the following final result,
\be
H_{10}(t_i,t_j)=\frac{t_i}{t_j}\e^{-t_j}\sum_{l=0}^{N-1}\frac{1}{(l!)^2}\tL_l(t_i)\tL_l(t_j)\ ,
\label{H10final}
\ee
which is proportional to the kernel of ordinary Laguerre polynomials~(\ref{normLag}). The remaining kernel $H_{11}$ can be expressed in terms of the two integral transforms which we have already computed,
\be
H_{11}(t,s)=\sum_{l=0}^{N-1}\frac{1}{h_l^{(M)}}\psi_l^{(M)}(t)\chi_l^{(M)}(s)
-\ G^{M-1,\,0}_{0,\,M-1}\left(\mbox{}_{0,\ldots,0}^{-} \bigg| \, \frac{s}{t}\right)\ .
\label{H11final}
\ee
This completes the computation of all $(k,l)$-point correlation functions in the 2mm, together with eq. (\ref{Rkldet}).

Although we will postpone the detailed analysis of the large-$N$ limit to future work let us mention the following nontrivial identity with respect to the kernel $H_{11}$:
\be
\sum_{l=0}^{\infty}\frac{1}{(l!)^2}t\tL_l(t)\chi_l^{(M)}(s)
=G^{M-1,\,0}_{0,\,M-1}\left(\mbox{}_{0,\ldots,0}^{-} \bigg| \, \frac{s}{t}\right)\ ,
\label{Gsumid}
\ee
implying that $\lim_{N\to\infty}H_{11}(t,s)=0$.
Assuming that the sum converges and can be integrated piecewise this can be shown as follows. In appendix \ref{OPmom1} we verify that $p_j^{(M)}(s)$ and $\chi_l^{(M)}(s)$ form a set of orthogonal functions with respect to the flat measure, see eq. (\ref{biOF}).
After multiplying both sides of eq. (\ref{Gsumid}) with $p_j^{(M)}(s)$ and integrating $s$ over $\mathbb{R}_+$ we obtain
\be
(j!)^{M-1}t\tL_j(t)=\int_0^\infty ds\ p_j^{(M)}(s)G^{M-1,\,0}_{0,\,M-1}\left(\mbox{}_{0,\ldots,0}^{-} \bigg| \, \frac{s}{t}\right)=\psi_j^{(M)}(t)\ ,
\ee
which is consistent with eq. (\ref{psidef}).

The identity (\ref{Gsumid}) most likely implies that in the naive large-$N$ limit, meaning $s$, $t$ and $M$ fixed, the correlation function $R_{k,l}^{(M)}$ factorises into $s$- and $t$-dependent parts regardless if $H_{00}$ vanishes or not since the determinant~\eqref{Rkldet} factorises into a $k\times k$ determinant incorporating $H_{01}$ and a $l\times l$ determinant comprising the block $H_{10}$. Therefore the correlation functions of the singular values $t_j$ of the matrix $X_1$ decouple and become the ones of the standard Wishart-Laguerre type.
There may be other ways to obtain a nontrivial coupling between singular values $s_a$ and $t_j$ in a more sophisticated large-$N$ limit (like in the so-called weak limit in \cite{ADOS}).

We finally make contact again to the one-matrix model formulation~(\ref{Zev1MM}).
The following ortho\-gonality relation which follows from eq.~\eqref{biOPdef} is explicitly verified in appendix \ref{OPmom1}
\be
\int_0^\infty ds\,p_i^{(M)}(s)\, \chi_j^{(M)}(s)=(i!)^{M+1}\delta_{ij}\ ,
\label{biOF}
\ee
in other words
$p_n^{(M)}(s)$ and $\chi_l^{(M)}(s)$ constitute a set of biorthogonal functions with respect to a single variable with flat measure.
In particular this relation  results into the following property of the kernel $H_{01}$, see eq. (\ref{H01pchi}), that contains the two:
\be
\int_0^\infty ds'H_{01}(s,s')H_{01}(s',s'')=H_{01}(s,s'')\ .\label{3.35}
\ee
We have thus closed the circle back to the jpdf~(\ref{jpdf1MM}) where we could directly replace
\bea
\Delta_N(s)
\det_{1\leq c,d\leq N}\left[
G^{M,\,0}_{0,\,M}\left(\mbox{}_{0,\ldots,0,d-1}^{-} \bigg| \, s_c\right)\right]&=&
\det_{1\leq a,b\leq N}[p_{a-1}^{(M)}(s_b)]\det_{1\leq c,d\leq N}[\chi_{c-1}^{(M)}(s_d)]\nn\\
&=& \prod_{j=0}^{N-1}h_j^{(M)}\det_{1\leq a,b\leq N}\left[H_{01}(s_a,s_b)
\right]\ .
\eea
This uses the invariance property of determinants under addition of columns, and then proceeds with standard techniques to deduce the correlation functions. This directly leads from eq.~(\ref{jpdf1MM}) to eq.~(\ref{1MMdet}) for the jpdf. With the property~\eqref{3.35} of the kernel we deduce eq. (\ref{Rkfinal}) from Dyson's theorem \cite{Mehta}.

\subsection{Spectral density, its moments and large-$N$ scaling
}\label{R1lim}

In this subsection we discuss in more detail the implications of our results for the spectral density of singular values.
Starting from the expression eq. (\ref{density}) which we repeat here in two equivalent forms,
\bea
R_{1}^{(M)}(s)
&=&\sum_{l=0}^{N-1}
\sum_{i,j=0}^l\frac{(-1)^{j+i}(l!)^2}{(l-j)!(l-i)!(i!)^2(j!)^{M+1}}\
G^{M,\,0}_{0,\,M}\left(\mbox{}_{j,\ldots,j,i+j}^{-} \bigg| \, {s}\right)
\nn\\
&=&\sum_{j=0}^{N-1}
\frac{1}{j!} ~_1F_{M}(-j;1,\ldots,1;s)
G^{M,\,1}_{1,\,M+1}\left(\mbox{}_{0,\ldots,0}^{-j} \bigg| \, s\right)
\ ,
\nn
\eea
we can explicitly compute expectation values for the moments for finite-$N$.
Starting from the first expression we obtain
\bea
\mathbb{E}[s^k]&\equiv&\frac{1}{N}\int_0^\infty ds\ s^k R_{1}^{(M)}(s)
\label{vevdef}\\
&=& \frac{1}{N}\sum_{l=0}^{N-1}
\sum_{i,j=0}^l\frac{(-1)^{j+i}(l!)^2(i+j+k)!((j+k)!)^{M-1}}{(l-j)!(l-i)!(i!)^2(j!)^{M+1}}\ .
\eea
Here we have normalised by eq. (\ref{R1norm}) and we have used again the identity eq. (\ref{Gmomid}) for moments of the Meijer $G$-function.
On the other hand using the compact expression for $\chi_j^{(M)}(s)$ in the second formulation of the density we can obtain a more concise result in the following way:
\bea
\mathbb{E}[s^k]
&=&\frac{1}{N}
\sum\limits_{j=0}^{N-1}\frac{1}{h_j^{(M)}}
\sum\limits_{l=0}^j\frac{(-1)^{j-l}}{(j-l)!}\left(\frac{j!}{l!}\right)^{M+1}
\int_0^\infty d s\,s^{l+k} (-1)^jG^{M,\,1}_{1,\,M+1}\left(\mbox{}_{0,\ldots,0}^{-j} \bigg| \, s\right)\nn\\
&=&\frac{1}{N}\sum\limits_{j=0}^{N-1}\sum\limits_{l=0}^j
\frac{(-1)^{j-l}}{(j-l)!(k+l-j)!}\left(\frac{(k+l)!}{l!}\right)^{M+1}\nn\\
&=&\frac{1}{N}\sum\limits_{l=0}^{N-1}\left(\frac{(k+l)!}{l!}\right)^{M+1}
\sum\limits_{j=0}^{N-1-l}\frac{(-1)^{j}}{j!(k-j)!}\nn\\
&=&\frac{1}{N}\sum\limits_{l=0}^{N-1}\left(\frac{(k+l)!}{l!}\right)^{M+1}
\int_0^{2\pi}\frac{d\varphi}{2\pi}\frac{1}{k!}(1-\e^{\imath\varphi})^k\frac{1-\e^{-\imath (N-l)\varphi}}{1-\e^{-\imath\varphi}}\nn\\
&=&\frac{1}{N}\sum\limits_{l=0}^{N-1}\frac{(-1)^{N-l-1}}{k!}\left(\frac{(k+l)!}{l!}\right)^{M+1}\left(\begin{array}{c} k-1 \\ N-l-1 \end{array}\right).
\label{momk}
\eea
We use the convention that inverse powers of factorials of negative integers give zero, rather than using the Gamma-function everywhere.
In the first step we have employed eq.~\eqref{Gmomidb}.
Interestingly,
the remaining sum can be further expressed in terms of a hypergeometric function if $k\geq N$,
\be
 \mathbb{E}[s^k]=(-1)^{N-1}\frac{(k!)^{M-1}(k-1)!}{N! \Gamma(k-N+1)}\ _{M+2}F_{M+1}(k+1,\ldots,k+1,1-N;1,\ldots,1,k-N+1;1)\ ,\ \
\ee
by extending the sum to infinity and comparing their Taylor series.
Indeed this relation can be generalised to $k<N$. Notice that in this case  the singular contributions in the hypergeometric function cancel with those in the Gamma-function in the denominator.

In the ensuing discussion we will need in particular the first moment $F_{N}^{(M)}$ for $k=1$ when rescaling the density, which can be readily read off
\be
\mathbb{E}[s]\equiv F_{N}^{(M)}=N^M\ .
\label{Fdef}
\ee
It agrees with the known case for $M=1$.
An alternative short derivation for the first moment is sketched in appendix \ref{OPmom1}.
Higher moments easily follow from eq.~\eqref{momk}, e.g. for the second moment we have
\be
\mathbb{E}[s^2]=\frac12 N^M\left((N+1)^{M+1}-(N-1)^{M+1}\right) \ .
\ee

\begin{figure}
\centering
\includegraphics[width=3in]{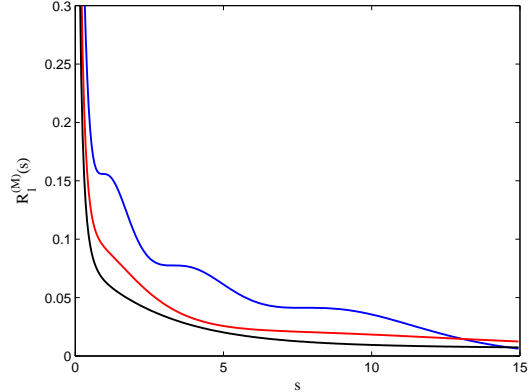}
\caption{Comparison of
the density eq. (\ref{density}) $R_{1}^{(M)}(s)$ for fixed $N=4$ without rescaling. The values $M=1,2,3$ correspond to the top (blue), middle (red) and bottom (black) curve, respectively.
}\label{fig:R1Ns}
\end{figure}

We illustrate our results for the density~(\ref{density}) by plotting it for various values of $N$ and $M$. As a first example in fig. \ref{fig:R1Ns} the density is shown for $M=1,2,3$ at fixed $N=4$. Clearly it is mandatory to know the right scale dependence of the correlation functions on $N$ and $M$. This means that after properly rescaling the bulk of the singular values is of order one, in order to be able to compare the density for finite $N$ at different values of $N$. In particular it is important to check the finite $N$-results against the limiting large-$N$ behaviour for different $M$, which has been derived for
products of quadratic \cite{BBCC,BG} and rectangular matrices \cite{2002Muller,BJLNS}.

Let us explain our procedure. First we normalise our density to unity, using eq. (\ref{R1norm}). Then we rescale the density by its first moment,
\be
\hat{R}_1^{(M)}(x)\equiv \frac{1}{N} F_{N}^{(M)} {R}_1^{(M)}\left(F_{N}^{(M)}x\right)\ ,
\label{rescale}
\ee
so that the new density $\hat{R}_1^{(M)}(x)$ has norm and first moment equal to unity\footnote{All densities satisfying this property will be denotes with a hat ``$\widehat{\quad}$''.}. Notice that this rescaling is
an alternative to 
the unfolding procedure onto the scale of the local mean level spacing. Instead of fixing the mean distance between two successive singular values, we fix here the singular values themselves, such that they are always of order one. This is exactly the macroscopic limit.

 Inserting eq. (\ref{Fdef}) we thus obtain a limiting density, which we denote by,
\be
\hat{\rho}^{(M)}(x)\equiv \lim_{N\to\infty}\hat{R}_1^{(M)}(x)=\lim_{N\to\infty} N^{M-1}{R}_1^{(M)}\left(N^M x\right)\ ,
\label{limdef}
\ee
that also has norm and first moment unity. Note that the known limiting density from the literature  may still have to be rescaled accordingly.
We can then compare $\hat{\rho}^{(M)}(x)$ and $\hat{R}_1^{(M)}(x)$ for different values of $N$ at fixed $M$.

\begin{figure}
\centering
\includegraphics[width=3in]{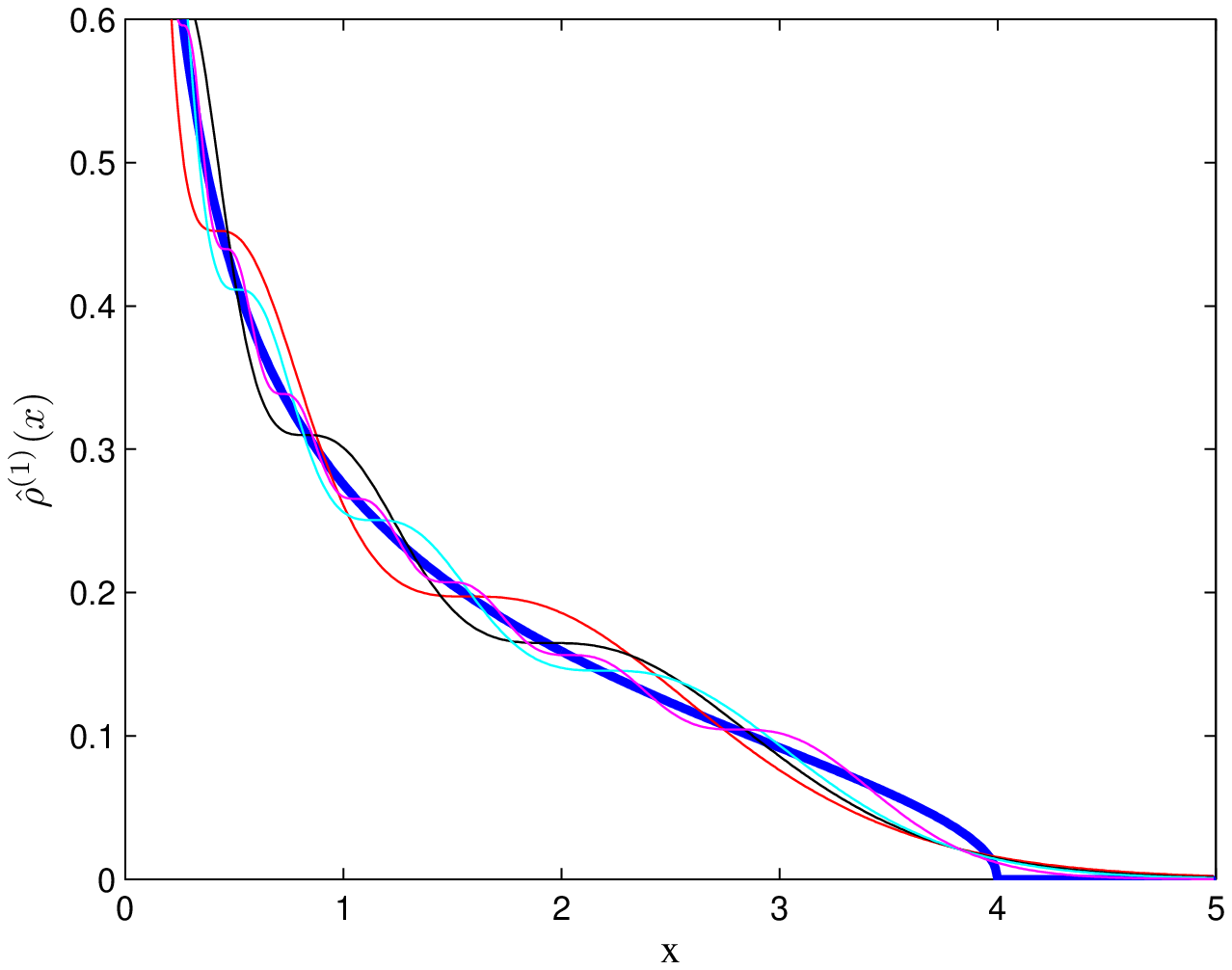}
\includegraphics[width=3in]{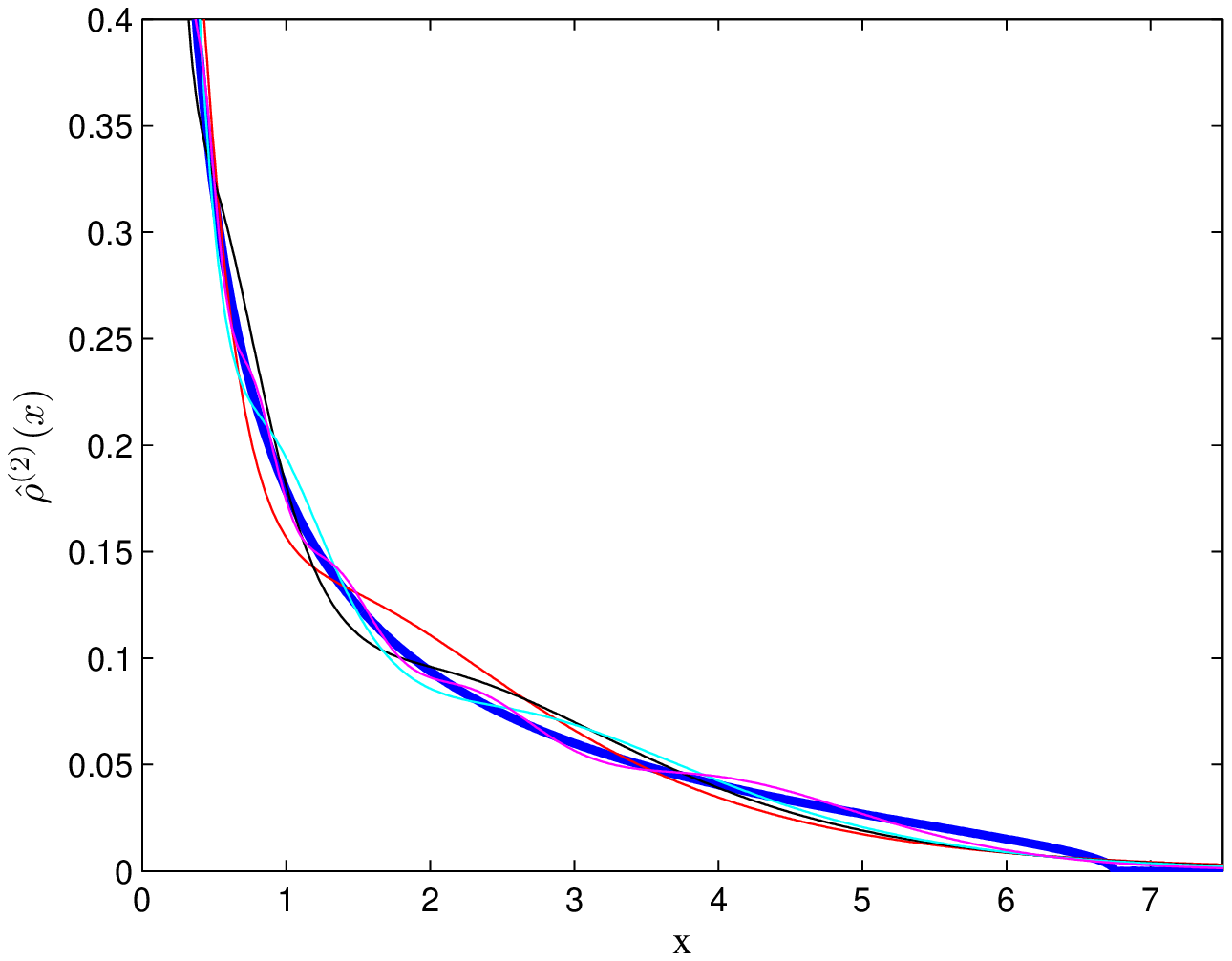}
\caption{
Left plot: the rescaled density $R_{1}^{(M=1)}(Nx)$ eq. (\ref{density}) for $M=1$ and $N=3,4,5,10$ (corresponding to red, black, cyan, and
magenta, respectively)
versus the limiting Marchenko-Pastur density eq. (\ref{MP}) (thick line).
Right plot:
the density eq. (\ref{density}) for $M=2$ rescaled as $N R_{1}^{(M=2)}(N^2 s)$, for $N=3,4,5,10$
compared to the limiting density $\hat{\rho}^{(M=2)}(x)$ (thick line).
}\label{fig:R1M1}
\end{figure}

This procedure can be illustrated with the simplest case $M=1$, the well known Wishart Laguerre ensemble. At large-$N$ we have
\be
\lim_{N\gg1}{R}_1^{(M=1)}(s)\approx \frac{1}{2\pi}\sqrt{\frac{4N-s}{s}}\Theta(4N-s),
\ee
which is the $N$-dependent Marchenko-Pastur density, and $\Theta(x)$ denotes the Heaviside function. With the first moment being given by $F_N^{(M=1)}=N$ we thus have
\be
\hat{\rho}^{(M=1)}(x)=\lim_{N\to\infty}N^0{R}_1^{(M=1)}\left(N x\right)=\frac{1}{2\pi}\sqrt{\frac{4-x}{x}}\ \Theta(4-x)\ ,
\label{MP}
\ee
for the rescaled and normalised density. This is the Marchenko-Pastur density with compact support on $(0,4]$. A comparison between $\hat{\rho}^{(M=1)}(x)$ and $\hat{R}_1^{(M=1)}(x)$ for various values of $N=3,4,5$ and 10 is given in fig. \ref{fig:R1M1} (left figure).

For $M>1$ the limiting expression for the density is not as explicit as in eq. (\ref{MP}). Here we will follow the notation of \cite{BJLNS} where a polynomial equation for the resolvent $G(z)$ was derived, which we display for the case of quadratic matrices only:
\be
\left(zG^{(M)}(z)\right)^{M+1}=z\left(zG^{(M)}(z)-1\right)\ .
\label{Gpoln}
\ee
The resolvent is related to the limiting spectral density by
\be
G^{(M)}(z)\equiv\int_0^\infty d\lambda \frac{\rho^{(M)}(\la)}{z-\la}\ ,
\ee
where $z$ is outside the support of the spectral density. This relation can be inverted as follows:
\be
\rho^{(M)}(\la)=-\frac1\pi \lim_{\epsilon\to0^+}\Im m\left(G^{(M)}(\la+i\epsilon)\right)\ .
\label{Ginversion}
\ee
For $M=1$ eq. (\ref{Gpoln}) reduces to a quadratic equation, which after taking the discontinuity along the support eq. (\ref{Ginversion}) leads to eq. (\ref{MP}), without further rescaling.

Increasing to $M=2$ the equation becomes cubic and we can still write out its solution,
which is chosen subject to boundary conditions and to yield a real density on the support $(0,3^3/2^2]$:\footnote{We thank Z. Burda for indicating how to determine the limiting support.}
\bea
G^{(M=2)}(z)&=&\frac{1}{\sqrt{3z}}\left(A_-^{-1/3}(z)+A_-^{1/3}(z)\right)=\frac{1}{\sqrt{3z}}\left((-A_+(z))^{1/3}+A_-^{1/3}(z)\right)\ ,\nn\\
A_{\pm}(z)&\equiv& \sqrt{\frac{27}{4z}-1}\pm\sqrt{\frac{27}{4z}}\ .
\label{Gcubic}
\eea
The density $\hat{\rho}^{(M=2)}(s)$ that is obtained from eq. (\ref{Gcubic}) by taking the discontinuity according to eq. (\ref{Ginversion}) (which happens to have the first moment equal to unity without further rescaling) is shown in fig. \ref{fig:R1M1} (right figure) in comparison to our rescaled finite-$N$ result $\hat{R}_{1}^{(M=2)}(s)$ for various values of $N$. As in the known case $M=1$ we obtain a nice agreement for $M=2$. An alternative derivation of $\hat{\rho}^{(M=2)}(s)$ via multiple orthogonal polynomials was recently presented in Refs.~\cite{Karol,Zhang}, with which our density agrees.

As a final remark for larger $M$ it might be useful to resolve the singularity of the density at the origin \cite{BJLNS},
$\lim_{s\to0}\hat{\rho}^{(M)}(s)\sim s^{-M/(M+1)}$,
by changing variables, just as it is well known that for $M=1$ a change to squared variables maps the Marchenko-Pastur density eq. (\ref{MP}) to the semi-circle.
A more detailed comparison to existing large-$N$ results for the density, its moments and support would require a careful asymptotic analysis of the special functions constituting our finite-$N$ density, and is postponed to future work.

\sect{Application to telecommunication}\label{app}

Consider a MIMO network with a single source and destination, both equipped with $N$ antennas. Information transmitted by the source is conveyed to the destination via $M-1$ successive clusters of scatterers, where each cluster (layer) is assumed to have $N$ scattering objects. Such a channel model proposed in \cite{2002Muller} is typical in modelling the indoor propagation of information between different floors \cite{2007Saunders}.

We assume that the vector-valued transmitted signal propagates from the transmitter array to the first cluster, from the first to the second cluster, and so on, until it is received from the $(M-1)$-st cluster by the receiver antenna array. Each communication channel is described by a random complex Gaussian matrix, and as a result the effective channel of this multi-layered model equals the product matrix $P_M$, see eq. (\ref{PMdef}).

\begin{figure}
\centering
\includegraphics[width=5in]{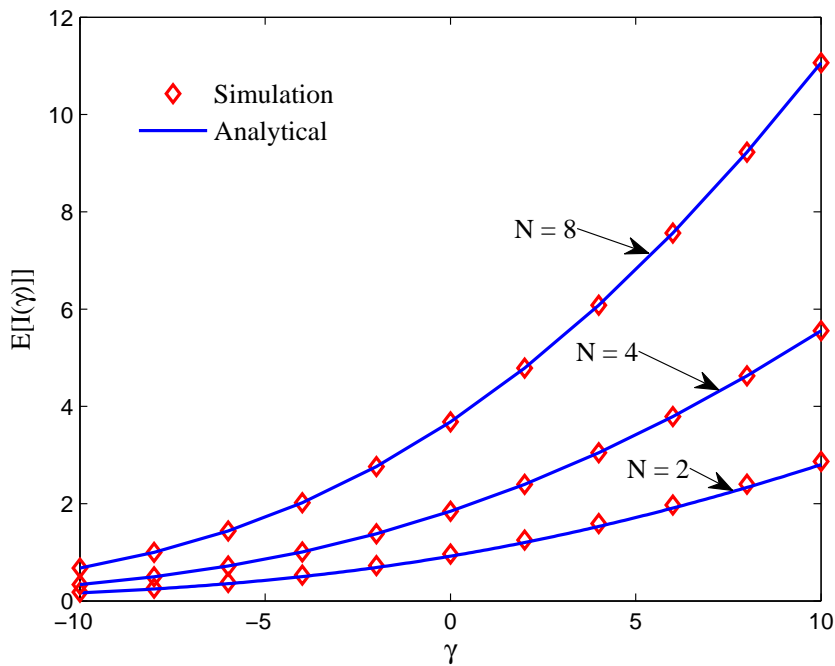}
\caption{The ergodic mutual information of multi-layered scattering MIMO channels with a fixed number of $2$ clusters ($M=3$) with a different number of scatters $N=2,4,8$.}\label{fig:M3}
\end{figure}

For the described communication channels, the mutual information measured in units of the natural logarithm (nats) per second per Hertz is defined as
\begin{equation}
\label{eq:MI}
\mathcal{I}(\gamma)\equiv\ln\det\left(I_{N}+\frac{\gamma}{N^M}P_{M}P_{M}^{\dag}\right)=\sum_{a=1}^{N}\ln\left(1+\frac{\gamma}{N^M}s_{a}\right),
\end{equation}
where $\gamma$ defines the average received signal-to-noise ratio per antenna which is a constant.
We employ the distribution~ (\ref{density}) of squared singular values
to compute its average.
The quantity of interest is called the ergodic mutual information of such channels. It is given by the expectation value of the random variable $\mathcal{I}(\gamma)$. Using the analogue of the expression~(\ref{vevdef}) from the previous section  we have
\begin{eqnarray}
\mathbb{E}\left[\mathcal{I}(\gamma)\right] &=& N\mathbb{E}\left[\ln\left(1+\frac{\gamma}{N^M}s\right)\right]
\nn\\
&=& \sum_{l=0}^{N-1}\sum_{i,j=0}^l\frac{(-1)^{i+j}(l!)^2}{(l-j)!(l-i)!(i!)^2(j!)^{M+1}}
\int_{0}^{\infty} ds~G^{M,\,0}_{0,\,M}\left( \begin{array}{c}-\\j,\ldots,j,i+j\\\end{array} \bigg| \, {s}\right) \ln\left(1+\frac{\gamma}{N^M}s\right)
\nn\\
&=&
\sum_{l=0}^{N-1}\sum_{i,j=0}^l\frac{(-1)^{i+j}(l!)^2}{(l-j)!(l-i)!(i!)^2(j!)^{M+1}}
G^{M+2,\,1}_{2,\,M+2}\left( \begin{array}{c}0,1\\0,0,j+1,\ldots,j+1,i+j+1\\\end{array} \bigg| \, {\frac{N^M}{\gamma}}\right)\!. \ \ \ \ \ \
\label{eq:EMI}
\end{eqnarray}
The last step is obtained by first replacing the logarithm by a Meijer $G$-function, eq. (\ref{Gid3}), and then applying the integral identity~(\ref{Gintid2}) from the appendix.
Note that the corresponding ergodic mutual information for the traditional MIMO channel
model, i.e. $M=1$, was derived in~\cite{1999Telatar}.

In order to get an independent confirmation of our analytical result~(\ref{eq:EMI}) we compare it to numerical simulations as follows.
We plot the ergodic mutual information~(\ref{eq:EMI}) in fig.~\ref{fig:M3} against Monte-Carlo simulations as a function of $\gamma$ in decibel (dB) for a given number of clusters, $M-1$, as an example, with different numbers of scatters per cluster, $N$.
Each simulated curve is obtained by averaging over $10^{6}$ independent
realisations of $P_M$. The statistical error bar is smaller than the symbol for the simulation in our plots.
The comparison in fig.~\ref{fig:M3} shows a $2$-layered scattering channel, i.e. $M=3$, with the number of scatters per layer varying from  $N=2,\ 4,\ \text{to}\ 8$.
We have also compared our results to simulations for other values of $N$ and $M$.

\sect{Conclusions and open questions}\label{conc}

In this paper we have derived the joint probability distribution function (jpdf) of singular values for any finite product of $M$ quadratic random matrices of finite size $N\times N$, with complex elements distributed according to a Gaussian distribution. This generalises the Wishart-Laguerre (also called chiral Gaussian) Unitary Ensemble which we recover for $M=1$. Starting from the jpdf we have computed all $k$-point density correlation functions of the singular values, by taking a detour over a two-matrix model like representation of the same model. In that way we showed that the jpdf being proportional to a Vandermonde times the determinant of Meijer $G$-functions
represents a determinantal point process. Its kernel of orthogonal functions generates all $k$-point functions in the standard way using Dyson's theorem. We also solved the auxiliary two-matrix model that couples a single matrix to the product of $M$ matrices, by constructing the biorthogonal polynomials explicitly, as well as the corresponding four kernels with their integral transforms.
On the way we found some nontrivial identities and integral representations of the Meijer $G$-function.

The density of the singular values are discussed in more detail at finite $N$ and $M$, including all its moments. We identified the macroscopic scaling to match the density with the known large-$N$ results for the macroscopic density of singular values. As a further application we have computed the averaged mutual information for multi-layered scattering of MIMO channels and have compared them to Monte-Carlo simulations for small $M$ and $N$.

Previous results for the macroscopic large-$N$ density of the singular values of quadratic or rectangular matrices and its expectation values of traces have mainly been obtained from probabilistic methods, in particular using free random variables.
The explicit results that we have obtained for the jpdf and all correlation functions thereof open up the possibility of another direction. One can now investigate the microscopic scaling limits zooming into various parts of the spectrum, by performing the asymptotic analysis of the orthogonal polynomials and their integral transforms that we computed.
Since the ensemble represents a determinantal point process one can also investigate the limiting distribution of individual eigenvalues, e.g. by considering their Fredholm determinant representation.
Moreover, one can now study the distributions of linear statistics such as the trace of the derived ensemble as well.

Based on known results for the universality of the spectrum of random matrices we expect the following outcome of such an analysis. The bulk and the soft edge behaviour of the spectrum should be governed by the universal Sine- and Airy-kernel, respectively, after unfolding and scaling appropriately. This includes that at the soft edge we expect to find the Tracy-Widom distribution, and it will be very interesting to identify the right scaling for that.
In contrast at the hard edge we expect to find new universality classes labelled by $M$. Already the way the macroscopic density diverges at the origin depends on it. This would fit into what was found recently for the complex eigenvalue spectrum of products of independent Ginibre matrices.

Other open problems include a generalisation of our construction to products of rectangular matrices, which seems quite feasible. Also the inclusion of determinants (or characteristic polynomials) into the weight, which should be related to rectangular matrices, seems within reach.
On the other hand going to non-Gaussian weight functions or investigating other symmetry classes with real or quaternion real matrix elements are very challenging problems. The reason is that our method depends crucially on
the Harish-Chandra--Itzykson-Zuber group integral to eliminate the angular variables. For the other symmetry classes no such explicit tool is presently at hand.
\\[2ex]

{\bf Acknowledgments:}
We would like to thank Zdzis{\l}aw Burda and Ralf M\"uller for useful discussions and correspondence.
The Laboratori Nazionali di Frascati are thanked (G.A.) for their hospitality where
part of this work was done.
We acknowledge partial support by the SFB $|$ TR12 ``Symmetries and Universality
in Mesoscopic Systems'' of the German research council DFG (G.A.), by the Alexander von Humboldt foundation (M.K.),  as well as by the Finnish Centre of Excellence in Computational Inference Research and the Nokia Foundation (L.W.).

\begin{appendix}

\sect{Some integral identities for Meijer $G$-functions}\label{G-Id}

In this appendix we collect a few integral representations and identities for the so-called Meijer $G$-function. It is defined as \cite{Gradshteyn}
\begin{equation}
G_{p,q}^{m,n}\left(\begin{array}{cccc}                                                                a_1, & a_2, & \ldots, & a_p \\
b_1, & b_2, & \ldots, & b_q
\end{array}\biggl|z
\right)=\frac{1}{2\pi i}\int\limits_{\cal C}du\,\frac{\prod_{j=1}^m\Gamma(b_j-u)\prod_{j=1}^n\Gamma(1-a_j+u)}
{\prod_{j=m+1}^q\Gamma(1-b_j+u)\prod_{j=n+1}^p\Gamma(a_j-u)}\,z^u\ .
\label{Gint1}
\end{equation}
The contour of integration ${\cal C}$ goes from $-i\infty$ to $+i\infty$ such that all poles
of the Gamma functions related to the $b_j$
lie to the right of the path, and all poles related to the $a_j$  to the left of the path. We are in particular interested in the case
\be
G^{m,\,0}_{0,\,m}\left(\mbox{}_{b_1,\ldots,b_m}^{-} \bigg| \, z\right)
=\frac{1}{2\pi i}\int\limits_{\cal C}du\,z^u \prod_{j=1}^m\Gamma(b_j-u)\ .
\label{Gint2}
\ee
Note that the function is symmetric in all its indices $b_1,\ldots,b_m$.
Special cases for small $m$ are given by \cite{Gradshteyn}
\bea
G^{1,\,0}_{0,\,1}\left(\mbox{}_{b_1}^{-} \bigg| \, z\right)&=&z^{b_1}\e^{-z} \ ,
\label{Gid1}\\
G^{2,\,0}_{0,\,2}\left(\mbox{}_{b_1,b_2}^{-} \bigg| \, z\right)&=&2z^{(b_1+b_2)/2}K_{b_1-b_2}(2\sqrt{z})\ ,
\label{Gid2}
\eea
and \cite{1990Prudnikov}
\bea
G^{1,\,2}_{2,\,2}\left(\begin{array}{cc}1&1\\1&0\\ \end{array} \bigg| \, z\right)&=&\ln(1+z)\ ,
\label{Gid3}
\eea
where $K$ is the modified Bessel function of the second kind equivalently known as the Macdonald function. It is possible to absorb powers of the argument of the Meijer $G$-function, due to the following shift \cite{Gradshteyn}
\be
z^k G_{p,q}^{m,n}\left(\begin{array}{ccc}
a_1, & \ldots, & a_p \\
b_1, & \ldots, & b_q
\end{array}\biggl|z
\right)=
G_{p,q}^{m,n}\left(\begin{array}{ccc}
a_1+k,  & \ldots, & a_p+k \\
b_1+k, & \ldots, & b_q+k
\end{array}\biggl|z
\right).
\label{shift}
\ee

The integral identities we collect here are for the moments of the Meijer $G$-function:
\bea
\int_0^\infty dt\,t^{n-1}
G^{m,\,0}_{0,\,m}\left(\mbox{}_{b_1,\ldots,b_m}^{-} \bigg| \, t\right)
&=&\prod_{j=1}^m\Gamma(b_j+n)\ ,
\label{Gmomid}\\
\int_0^\infty d s\,s^{n} G^{M,\,1}_{1,\,M+1}\left(\mbox{}_{0,\ldots,0}^{-j} \bigg| \, s\right)&=&\underset{\varepsilon\to0}{\lim}\frac{(n!)^M\Gamma(j-n+\varepsilon)}{\Gamma(-n+\varepsilon)}=
(-1)^j\frac{(n!)^{M+1}}{\Gamma(n-j+1)}\ .
\label{Gmomidb}
\eea
We only state two particular cases, for the most general setting see e.g. \cite{1990Prudnikov}. In the last step of the second identity we have employed the recursion of the Gamma-function, $\Gamma(j-n+\varepsilon)=$\newline$\Gamma(-n+\varepsilon)\prod_{l=0}^{j-1}(l-n+\varepsilon)$.

The second integral identity needed in section \ref{jpdf} follows easily from the representation eq. (\ref{Gint2}):
\bea
&&\int_0^\infty dt\,t^{d-2}\e^{-t}
G^{m,\,0}_{0,\,m}\left(\mbox{}_{b_1,\ldots,b_m}^{-} \bigg| \, \frac{s}{t}\right)
=\int_0^\infty \frac{dv}{v} \left(\frac{s}{v}\right)^{d-1}\e^{-s/v}
G^{m,\,0}_{0,\,m}\left(\mbox{}_{b_1,\ldots,b_m}^{-} \bigg| \, v \right)\nn\\
&=&\frac{1}{2\pi i}\int\limits_{\cal C}du \prod_{j=1}^m\Gamma(b_j-u)
\int_0^\infty dt\,t^{d-2}\e^{-t}\left(\frac{s}{t}\right)^u
=\frac{1}{2\pi i}\int\limits_{\cal C}du \prod_{j=1}^m\Gamma(b_j-u)\, s^u \Gamma(d-1-u)\nn\\
&=&G^{m+1,\,0}_{0,\,m+1}\left(\mbox{}_{b_1,\ldots,b_m,d-1}^{-} \bigg| \, s\right)\ .
\label{Gshiftid}
\eea
In the first line we simply substituted $t\to v=s/t$ to obtain a second version of the identity. The rest follows from the definition of the Meijer $G$- and the Gamma-function. Notice that for $d=1$ and $b_1=\ldots=b_m=0$ this recursion for the Meijer $G$-function was already derived in \cite{ABu}.

We are now prepared to show the multiple integral representation of the Meijer $G$-function that we need in the derivation of the jpdf in section \ref{jpdf}. It slightly generalises the representation found in \cite{ABu}. The statement is that
\be
G^{m,\,0}_{0,\,m}\left(\mbox{}_{0,\ldots,0,b}^{-} \bigg| \, \frac{x_m}{x_0}\right)=\int_0^\infty\frac{dx_1}{x_1} \left(\frac{x_1}{x_0}\right)^{b}
\int_0^\infty\frac{dx_2}{x_2}\ldots \int_0^\infty\frac{dx_{m-1}}{x_{m-1}}\prod_{j=1}^m\e^{-x_j/x_{j-1}}\ ,
\label{Gintid}
\ee
for $m>1$. Comparing this equation to eq. (\ref{1MMreduct})  with $b=d-1$, we work with $m=M$ squared singular values, $x_j=(\laj_c)^2$, where we introduce a dummy variable $x_0$. The variable $x_0$ will be set to unity for our original purposes. However, it will be useful when applying the identity to eq. (\ref{2MMreduct}). Our proof goes by induction in $m$. For $m=2$ \cite{Gradshteyn} we have
\be
\int_0^\infty\frac{dx_1}{x_1} \left(\frac{x_1}{x_0}\right)^{b}\e^{-\frac{x_1}{x_0}-\frac{x_2}{x_1}}
=2\left(\frac{x_2}{x_0}\right)^{b/2}K_{-b}\left(2\,\sqrt{{x_2}/{x_0}}\right)=
G^{2,\,0}_{0,\,2}\left(\mbox{}_{0,b}^{-} \bigg| \, \frac{x_2}{x_0}\right),
\ee
where the last step is due to eq. (\ref{Gid2}). This leads to
\be
\int_0^\infty\frac{dx_2}{x_2}G^{2,\,0}_{0,\,2}\left(\mbox{}_{0,b}^{-} \bigg| \, \frac{x_2}{x_0}\right)\e^{-x_3/x_2}=G^{3,\,0}_{0,\,3}\left(\mbox{}_{0,b,0}^{-} \bigg| \, \frac{x_3}{x_0}\right)\ ,
\ee
for $m=3$, where we have applied the identity eq. (\ref{Gshiftid}) in its second form shown in the first line, with $d=1$ and $v=x_2/x_0$. For the induction step $m-1\to m$ we simply have to repeat the same procedure, which follows easily from the very same identity
\be
\int_0^\infty\frac{dx_m}{x_m}G^{m,\,0}_{0,\,m}\left(\mbox{}_{0,\ldots,0,b}^{-} \bigg| \, \frac{x_m}{x_0}\right)\e^{-x_{m+1}/x_{m}}=G^{m+1,\,0}_{0,\,m+1}\left(\mbox{}_{0,\ldots,0,b}^{-} \bigg| \, \frac{x_{m+1}}{x_0}\right)\ ,\label{a12}
\ee
which completes the proof.

Note that the same identity (\ref{Gintid}) can be used to provide the second step in eq. (\ref{2MMreduct}), when setting $b=0$, $m=M-1$ and shifting the indices of the variables $x_{j-1}=(\laj_c)^2$ for $j=1,\ldots,M$. This is the reason why $x_0$ is useful.

Finally we state an integral identity needed in section \ref{app}, concerning the integral of two Meijer $G$-functions,
\be
\int_0^\infty ds\ G^{1,\,2}_{2,\,2}\left(\begin{array}{cc}1&1\\1&0\\ \end{array} \bigg| \, s\frac{\gamma}{N}\right)
G^{M,\,0}_{0,\,M}\left( \begin{array}{c}-\\j,\ldots,j,i+j\\\end{array} \bigg| \, {s}\right)
=\frac{N}{\gamma}
G^{M+2,\,1}_{2,\,M+2}\left( \begin{array}{c}-1,0\\-1,-1,j,\ldots,j,i+j\\\end{array} \bigg| \, {\frac{N}{\gamma}}\right).\ \ \ \
\label{Gintid2}
\ee
Notice that it is a particular choice of a general formula \cite{1990Prudnikov}. In order to arrive at eq. (\ref{eq:EMI}) we apply the shift (\ref{shift}).

\sect{Orthogonality check and first moment}\label{OPmom1}

In this appendix we explicitly confirm both the orthogonality~(\ref{biOPdef}) of the bOP $p_n^{(M)}(s)$ and $q_l^{(M)}(t)$ with respect to two variables, as well as the fact that
$p_n^{(M)}(s)$ and $\chi_l^{(M)}(s)$ constitute a set of biorthogonal functions with respect to one variable with flat measure, eq. (\ref{biOF}).
Although being true by construction we will see the orthogonality ultimately boils down to the standard orthogonality of Laguerre polynomials.

The biorthogonal polynomials that were constructed in subsection \ref{bOP} using the bimoment matrix
must automatically satisfy the orthogonality relation (\ref{biOPdef}).
We will check this here independently, which
implies at the same time that one of the polynomials and the integral transform~(\ref{chijdef}) of the other are orthogonal functions (as they should be, in order to constitute proper kernels):
\bea
&&\int_0^\infty ds\,p_i^{(M)}(s)\, \chi_j^{(M)}(s)=\int_0^\infty ds\int_0^\infty dt\, w^{(M)}(s,t)\,p_i^{(M)}(s) q_j^{(M)}(t)
\nn\\
&=&\int_0^\infty dt\,\e^{-t}\tL_j(t)
\sum_{k=0}^i\frac{(-1)^{i-k}}{(i-k)!}\left(\frac{i!}{k!}\right)^{M+1}
\int_0^\infty \frac{ds}{t}
G^{M-1,\,0}_{0,\,M-1}\left(\mbox{}_{0,\ldots,0}^{-} \bigg| \, \frac{s}{t}\right)
s^k\nn\\
&=&\int_0^\infty dt\,\e^{-t}\tL_j(t)
\sum_{k=0}^i\frac{(-1)^{i-k}}{(i-k)!}\left(\frac{i!}{k!}\right)^{M+1} t^k (k!)^{M-1}
=(i!)^{M-1}\int_0^\infty dt\,\e^{-t}\tL_j(t)\tL_i(t)\nn\\
&=&(i!)^{M+1}\delta_{ij}\ .
\label{biOPcheck}
\eea
Here we have used the identity~\eqref{Gmomid} for moments of the Meijer $G$-function, which cancels the extra factorials in the generalised Laguerre polynomials $p_i^{(M)}(s)$, see eq. (\ref{pnres}) after the first integration over $s'=s/t$. The last step follows from the known orthogonality of Laguerre type.

In the second part of this appendix we provide a much simpler, probabilistic argument that leads to the first moment eq. (\ref{Fdef}). Noting that
\be
\mathbb{E}\left[X_jX_j^\dag\right]=NI_N\ ,\ \ \mbox{for}\ \ j=1,\ldots,L
\ee
we have that
\bea
\mathbb{E}[s]&=&\frac1N \mathbb{E}\left[\sum_{a=1}^Ns_a\right]=\frac1N \mathbb{E}\left[\Tr(P_MP_M^\dag)\right]\nn\\
&=&\frac1N \Tr\left(\prod_{j=1}^M\mathbb{E}\left[X_jX_j^\dag\right]\right)=
\frac1N N^M\Tr(I_N)=N^M\ ,
\eea
where we have reordered successively the $X_j$ under the trace in the first line.

\end{appendix}


\end{document}